\documentclass[letterpaper, 10 pt, conference]{ieeeconf}  

\IEEEoverridecommandlockouts                              
\def\BibTeX{{\rm B\kern-.05em{\sc i\kern-.025em b}\kern-.08em
    T\kern-.1667em\lower.7ex\hbox{E}\kern-.125emX}}

\usepackage{mathtools}
\usepackage{mathbbol}
\usepackage{graphicx} 
\usepackage{amsfonts}
\usepackage{amsmath} 
\usepackage{amssymb}  
\usepackage{bbm}

\newtheorem{theorem}{Theorem}

\newtheorem{lemma}[theorem]{Lemma}
\newtheorem{corollary}[theorem]{Corollary}

\title{\LARGE \bf
Designing Laplacian flows for opinion clustering in structurally balanced and unbalanced networks
}

\author{$^{1}$Vishnudatta Thota,  $^{2}$Twinkle Tripathy and $^{3}$Debasattam Pal
\thanks{$^{1}$Vishnudatta Thota is a Project Associate and $^{2}$Twinkle Tripathy is an Assistant Professor in the Department of Electrical Engineering, Indian Institute of Technology Kanpur, Uttar Pradesh, India 208016. Email:{ \tt\small thotav@iitk.ac.in, ttripathy@iitk.ac.in}
$^{3}$Debasattam Pal is an Associate Professor in the Department of Electrical Engineering, Indian Institute of Technology Bombay, Maharashtra 400076. Email:{ \tt\small
debasattam@ee.iitb.ac.in}
}%
}

\begin{document}

\maketitle
\thispagestyle{empty}
\pagestyle{empty}

\begin{abstract}
In this work, we consider a group of $n$ agents whose interactions can be represented using unsigned or signed structurally balanced graphs or a special case of structurally unbalanced graphs. A Laplacian-based model is proposed to govern the evolution of opinions. The objective of the paper is to analyze the proposed opinion model on the opinion evolution of the agents. Further, we also determine the conditions required to apply the proposed Laplacian-based opinion model. Finally, some numerical results are shown to validate these results.

\end{abstract}


\addtolength{\textheight}{-3cm}   


\section{INTRODUCTION}
Opinion dynamics is a field of study that examines the evolution and convergence of opinions of a group of interacting agents. 
The evolution of the opinions of the agents can lead to collective behaviour like consensus, polarization, and clustering of opinions within groups. 
To understand this collective behaviour, researchers have increasingly explored agent-based models, which has led to an increasing amount of literature on this subject.
These agent-based models are used  in various studies such as voting patterns (\cite{2014agent_based_model},\cite{negativebias2017}), trends in social networks (\cite{2011_social_network},
\cite{gorodnichenko2021social}), and collective animal behaviour 
(\cite{2008_animal_groups}, \cite{PRATT20051023}, \cite{srivastava2017bio}).

Various mathematical models have been proposed to explain the collective behaviour of opinion formation like consensus (\cite{Degroot1974}, \cite{ZHU2014552}, \cite{blondel_2009}, \cite{Qin_2017}), polarization (\cite{altafini_polarisation2012}, \cite{Song_2017}) and clustering (\cite{valcher2000algebraic}, \cite{Cont2007}, \cite{trust_mistrust2016}). Our focus in this paper is on the \textit{opinion clustering} behaviour of the evolution of agents, in which more than two clusters of agents' opinions are eventually formed in the network.
In \cite{altafini_polarisation2012}, the author defined the structural balance of a graph and explained how the agents are polarized by their proposed model. We will be using the same definition of structural balance and use our proposed model to show that the clustering behaviour is possible for structurally balanced graphs.
It is shown in \cite{trust_mistrust2016} that the clustering of opinions of the agents occurs using a DeGroot-based model when there is a subnetwork of structurally balanced and globally reachable nodes. 
These graphs are a special case of structurally unbalanced graphs, we will use our proposed model on these graphs to achieve desired clustering of opinions.
In \cite{consensus_cluster_valcher22}, the authors proposed a modified DeGroot model for clustering, wherein the agents were divided into clusters and the agents interact cooperatively within clusters and antagonistic between different clusters.
In \cite{ROY2015259}, the author defined the notion of scaled consensus in which the final opinion of the agents approaches a dictated ratio. In these works, the clustering of the final opinion states depends on the initial opinion states and model parameters, which are not the only factors that lead to the clustering of final opinion states.

In addition to inter-agent interactions, external influences can also impact the opinion evolution of the agents \cite{varma2020analysis}. For example, external influences such as news and social media are widely used to form opinions in socio-political scenarios. The effect of spreading misinformation and rumors on opinion formation is studied by the authors in \cite{vicaro2016misinfo}. In \cite{simple2019}, a simple model using a Monte Carlo approach was used to study how people develop opinions and vote in a two-party system, where the population is exposed to an external bias that benefits the minority.
The impact of exogenous influences is modeled as the stubbornness of the agent to its initial position in \cite{FRIEDKIN1997209}.

In contrast to the works discussed so far, the current work proposes the use of a modified out-degree matrix to achieve opinion clustering for unsigned and signed structurally balanced graphs. 
In this work, we show the existence of the proposed Laplacian-based model for the unsigned and signed structurally balanced graphs and a special case of structurally unbalanced graphs. 
Moreover, we also show that the proposed Laplacian-based model does not exist for an anti-balanced graph. The motivation for our work stems from the fact that the proposed Laplacian-based model can be used to mitigate undesirable effects like polarization and achieve opinion clustering.
The opinion evolution is governed by a Laplacian-based model. This modified Laplacian matrix can change the relative importance of an agent's opinion as compared to the opinion of its neighbors.
Thereafter, we show the final opinion states reached by following the proposed Laplacian-based model.
The major advantage of our model is that it is possible using this model, to reach different opinion clusters for unsigned and signed structurally balanced graphs.

The paper is organized as follows: Section \ref{Section-Preliminaries} contains some necessary preliminaries from graph theory. Section \ref{Sec:Opinion_modelling} presents the model which governs the evolution of opinions. The existence of the proposed Laplacian-based model is discussed in Section \ref{sec:Existence of the proposed out degree matrix}. The effect of the proposed Laplacian-based model on opinion formation is discussed in Section \ref{sec:Effect_of_proposed_out_degree}.
Section \ref{sec:Simulation results} demonstrates these results through numerical simulations. 
Section \ref{sec:Discussion}, discusses some of the improvements of our model as compared to the existing literature.  
Finally, Section \ref{Sec:conclusion and future works} concludes the paper with some insights into the possible future research directions.

\section{PRELIMINARIES}
\label{Section-Preliminaries}
A signed weighted graph is represented by $\mathcal{G}=(\mathcal{V},\mathcal{E},A)$ where $\mathcal{V}=\{1,2,\cdots,n\}$ is the set of nodes, $\mathcal{E} \subseteq \mathcal{V} \times \mathcal{V}$ is the set of edges, and $A$ is the adjacency matrix of the graph $\mathcal{G}$. 
The graph $\mathcal{G}$ is used to represent a multi-agent network whose agents are represented by the vertices of the graph $\mathcal{G}$ and the interactions between these agents are represented by the edges of the graph $\mathcal{G}$. In a multi-agent network, the interactions between the agents can be cooperative or antagonistic which is represented by the positive and negative sign of the weight of the edges of the graph $\mathcal{G}$  respectively.

The adjacency matrix $A \in \mathbb{R}^{n\times n}$ for the graph $\mathcal{G}$ is denoted by $A=\{a_{ij}\}$. The entry $a_{ij}$ equals the weight of the edge $(i, j)$ or is zero otherwise. 
If the adjacency matrix of a graph $\mathcal{G}$ is symmetric, then the graph is called undirected; if not, it is called a digraph.
The standard out-degree matrix for a signed digraph $\mathcal{G}$ is $D=diag\left(d_{i} \right)$ where $d_{i}=\sum_{k =1}^{n} |a_{ik}|$. $\mathbb{1}_n$ and $\mathbb{0}_n$ denote column vectors with all entries equal to $+1$ and $0$, respectively. The matrix $I_n$ denotes the $n \times n$ identity matrix.
 The standard Laplacian matrix $L \in   \mathbb{R}^{n\times n}$ for the unsigned graph $\mathcal{G}$ is defined as 
\begin{equation} \label{eq:laplacian}
L=D-A
\end{equation}
It follows from eqn. \eqref{eq:laplacian}, that $L\mathbb{1}_n=\mathbb{0}_n$ therefore, standard Laplacian matrix $L$ will always have a zero eigenvalue in a cooperative framework.
The non-zero eigenvalues of the Laplacian matrix have a strictly positive real part. 

The unsigned weighted graph $\Bar{\mathcal{G}}=(\mathcal{V},\mathcal{E})$ is defined for the corresponding signed weighted graph $\mathcal{G}$. 
The adjacency matrix $\Bar{A}$ of the graph $\Bar{\mathcal{G}}$ is denoted by  $\Bar{A}=\{|a_{ij}|\}$, where $a_{ij}$ equals the weight of the edge $(i, j)$ of the graph $\mathcal{G}$ or is zero otherwise. 
In a signed weighted graph, a negative cycle is characterized by the existence of at least one cycle in which the product of the edge weights is negative.
Next, we introduce some definitions of structural balance, anti-balance, and strict unbalance of the signed graph $\mathcal{G}$.

A signed graph $\mathcal{G}$ is said to be \textit{structurally balanced} if and only if there is a bipartition of the node set $\mathcal{V}$ into non-empty subsets  $\mathcal{V}_{1}$ and $\mathcal{V}_{2}$ such that $\mathcal{V} = \mathcal{V}_{1} \cup \mathcal{V}_{2}$, $\mathcal{V}_{1} \cap \mathcal{V}_{2}= \phi$ and any edge between the two node subsets is negative while any edge within each node subset is positive. A signed graph $\mathcal{G}$ is said to be \textit{structurally anti-balanced} if and only if there is a bipartition of the node set $\mathcal{V}$ into non-empty subsets  $\mathcal{V}_{1}$ and $\mathcal{V}_{1}$ such that $\mathcal{V} = \mathcal{V}_{1} \cup \mathcal{V}_{2}$, $\mathcal{V}_{1} \cap \mathcal{V}_{1}= \phi$ and any edge between the two node subsets is positive while any edge within each node subset is negative. A signed graph $\mathcal{G}$ is said to be \textit{structurally unbalanced} if $\mathcal{G}$ is neither structurally balanced nor structurally anti-balanced.

A set $\mathcal{K} \subset \mathbb{R}^{n} $ is called a cone if $\alpha\mathcal{K} \subset \mathcal{K} $ for all $\alpha \geq 0\} $. 
A cone is said to be solid if its interior is a non-empty set.
A cone is said to be closed if the limit points of converging sequences within the cone are also contained within it. 
A cone is said to be proper if it is closed, solid, convex ($\alpha x + \beta y \in \mathcal{K}$, given $x, y \in \mathcal{K}$ and $\alpha, \beta \geq 0$) and pointed ($\mathcal{K} \cap \{-\mathcal{K}\} = \{0\}$).
A cone $\mathcal{K}$ is said to be polyhedral if it can be expressed as the set of non-negative linear combinations of a finite set of generating vectors (extreme rays).
A matrix $\mathbf{C}_{n \times k}$ can be found such that $\mathcal{K}$ coincides with the set of non-negative combinations of the columns of $\mathbf{C}$.

\section{OPINION MODELLING}
\label{Sec:Opinion_modelling}
In this work, we consider a group of $n$ autonomous agents modeled as single integrators. The interactions among the agents are coopetitive (cooperative$/$competitive) in nature which makes the underlying network a signed graph $\mathcal{G}$. We study the evolution of the opinions of the agents in such networks. When the networks are large, clustering of opinions is a common phenomenon. For example in real-world scenarios like bimodal coalitions, duo-polistic markets, and competing international alliances. However, in general, the phenomenon of polarisation of opinions can be undesirable in a society in various scenarios e.g. communal riots. Hence, the objective of the paper is to propose an opinion model to achieve a \textit{a desired clustering} of opinions in the network. 

Let the opinions of the agents be represented by the vector $x=(x_1,x_2,\cdots,x_n)^T$ where $x_{i} \in \mathbb{R}$ represents the opinion of the $i^{th}$ agent. The Laplacian flow-based models have been used extensively in the literature to explain the behaviour of consensus in unsigned digraphs with globally reachable nodes. In \cite{altafini_polarisation2012}, Altafini showed a modified form of the Laplacian matrix for structurally balanced signed networks which results in bipartite consensus. The authors in papers (\cite{consensus_cluster_valcher22}, \cite{ROY2015259}) proposed more variants of the Laplacian flow-based models to achieve desired results like clustering. However, these works do not guarantee convergence to the desired values of the opinions i.e. a desired clustering of opinions.

In our work, we propose a modified Laplacian matrix ${L}_{x} = \theta_{x}-A$ where $L_{x}$ is the proposed Laplacian matrix of the graph $\mathcal{G}$, $\theta_{x}$ is the modified out-degree matrix of the graph $\mathcal{G}$, and $A$ is its standard adjacency matrix. The major difference between the proposed Laplacian matrix and the aforementioned ones is that we consider the stubbornness of the agents. 
The matrix $\theta_{x}$ of the proposed Laplacian-based model can be a non-diagonal matrix and the diagonal entries can be different from those of the standard out-degree matrix which is used to represent the stubbornness of the agents.  

Considering the proposed Laplacian matrix $L_{x}$, the opinion dynamics of the group of $n$ agents in vector form is,
\begin{equation} 
\label{eq:linear_system_1}
\dot{x}=-L_{x}x = -(\theta_{x}-A)x 
\end{equation}
where $x=(x_1,x_2,\cdots,x_n)^T$ and $x_{i}$ represents the opinion of the $i^{th}$ agent.

\section{DESIGN OF THE PROPOSED Laplacian Matrix}
\label{sec:Existence of the proposed out degree matrix}
In this section, we present a methodology to design the modified Laplacian matrix which guarantees a desired clustering of opinions in various graph structures. Given a graph structure, hence, its adjacency matrix, we know from eqn. \eqref{eq:linear_system_1} that designing a modified Laplacian matrix is equivalent to designing a suitable out-degree matrix $\theta_x$. 

To design a suitable $\theta_{x}$ for unsigned and signed structurally balanced graphs, we use an invertible matrix $P \in \mathbb{R}^{n\times n}$ to transform the coordinate system from $x$ to $z$,
\begin{equation} 
\label{eq:coordinate_change}
z=Px
\end{equation}
where $z=(z_1,z_2,\cdots,z_n)^T$ and $z_{i} \in \mathbb{R}$. Hence, the evolution of opinions can be described in the new coordinate system as,
\begin{equation} 
\label{eq:linear_system_2}
\dot{z}=-PL_{x}P^{-1}z  = -L_{z}z = -\left(\theta_{z} - A_{z}\right)z
\end{equation}
where $A_{z} =  PAP^{-1}$ is the adjacency matrix of the graph in the coordinate system $z$, $\theta_{z} = P\theta_{x} P^{-1}$ is the standard out-degree matrix computed using the adjacency matrix $A_{z}$ and $L_{z}$ is the Laplacian matrix in coordinate system $z$. Now, the procedure for designing $L_x$ is as given below:
\begin{itemize}
\item We define a set $\mathcal{P}$ as the set of all invertible matrices such that,
\begin{align}
\label{eq:A_condition}
\mathcal{P}:= \{P~|~det(P)\neq 0 \text{ and } PAP^{-1}\geq 0\}
\end{align}
In other words, any $P \in \mathcal{P}$ makes the adjacency matrix $A_{z}$ non-negative. 
\item The matrix $P \in \mathcal{P}$ is then used to find the standard out-degree matrix as, 
\begin{align}
\theta_{z} = diag\left(\mathbb{1}_n^{T}PAP^{-1} \right)
\end{align}
\item The modified out-degree matrix $\theta_{x}$ is then given by,
\begin{equation}
    \label{eq:proposed_outdegree}
    \theta_{x} =  P^{-1}\theta_{z} P = P^{-1}diag\left(\mathbb{1}_n^{T}PAP^{-1} \right) P
\end{equation}
The modified Laplacian matrix can be calculated using eqn. \eqref{eq:linear_system_1} as $L_x=\theta_x-A$.
\end{itemize}

Note that the above design procedure relies on the existence of a suitable $P\in\mathcal{P}$.
Therefore, a natural question is, does such a $P$ exist for any arbitrary graph structure and any desired value of the clustering vector? 
We pursue this question for a general class of unsigned and structurally balanced signed graphs. 

\begin{lemma}
\label{thm:unsigned_existence}
For any unsigned digraph $\mathcal{G}$, the set $\mathcal{P}$, defined in eqn. \eqref{eq:A_condition}, is always non-empty.
\end{lemma}
\begin{proof}
Since $\mathcal{G}$ is unsigned, the corresponding adjacency matrix $A$ is already non-negative. Consider an invertible matrix $P_{1}$ which is a diagonal matrix with positive diagonal entries. Then, $P_{1}^{-1}$ is also a diagonal non-negative matrix whose diagonal entries are positive. So $P_{1}AP_{1}^{-1} \geq 0$ which implies $P_{1} \in \mathcal{P}$. Hence, proved. 
\end{proof}

Note that any invertible diagonal matrix $P \geq 0$ can always satisfy the condition $PAP^{-1} \geq 0$ for unsigned graphs. Next, we pursue the same for signed graphs.

\begin{theorem}
\label{thm:signed_structurally_balanced_existence}
Consider a structurally balanced digraph $\mathcal{G}$ which has at least one negative cycle. When the unsigned graph corresponding to $\mathcal{G}$ is irreducible and aperiodic, the set $\mathcal{P}$ is always non-empty, where $\mathcal{P}$ is defined in eqn. \eqref{eq:A_condition}. 
\end{theorem}
\begin{proof}
It is known that the Perron-Frobenius theorem can be extended for structurally balanced signed digraphs to analyze their spectral properties \cite{tian2024spreading}. For such graphs, the maximum modulus eigenvalue of the adjacency matrix $A$ corresponding to graph $\mathcal{G}$ is positive and simple. Furthermore, the spectral radius of $A$ lies in its spectrum such that it is the largest in magnitude. 

We know from Theorem 3.1 in \cite{valcher2000algebraic} that the necessary condition for a matrix $B \in \mathbb{R}^{n\times n}$ to be positive is that $B$ can make a proper polyhedral cone $\mathcal{K} \in \mathbb{R}^{n}$ invariant. For the given signed digraphs, the adjacency matrix $A$ will leave a proper polyhedral cone $\mathcal{K} \in \mathbb{R}^{n}$ invariant as its spectral properties satisfy the conditions mentioned in Theorem 3.1 in \cite{valcher2000algebraic}. Furthermore, it also then satisfies all the necessary conditions to find an invertible matrix $P\in\mathcal{P}$ such that $PAP^{-1}$ is a non-negative matrix (Lemma 2.3 in \cite{valcher2000algebraic}). 

Given that the graph is structurally balanced, its vertices can be partitioned into two disjoined non-empty sets $\mathcal{V}_1$ and $\mathcal{V}_2$. Then, let us consider an invertible diagonal matrix $P_{2}$ whose diagonal entries corresponding to vertices in set $\mathcal{V}_1$ are positive and those in set $\mathcal{V}_2$ are negative. It will always be in set $\mathcal{P}$ because it can make $P_{2}AP_{2}^{-1}$ non-negative. Hence, proved.
\end{proof}

Thm. \ref{thm:signed_structurally_balanced_existence} shows the existence of a set of suitable matrices in $\mathcal{P}$ for a class of structurally balanced signed digraphs. Structurally balanced graphs are bipartite graphs as their vertices can be partitioned into two disjoint subsets. Similarly, in \textit{k-partite} graphs, the vertices can be partitioned into multiple disjoint subsets such that the intra-agent interactions are cooperative while the inter-group ones are antagonistic. For $n>2$, such graphs are generally structurally unbalanced.

\begin{corollary}
    Consider structurally unbalanced k-partite graphs. There does not exist a diagonal invertible matrix $P$ that can make $PAP^{-1}$ positive, where $A$ is the adjacency matrix of the k-partite graph and $\mathcal{P}$ is defined in eqn. \eqref{eq:A_condition}. 
\end{corollary}

\begin{proof}
    Consider a 3-partite graph whose vertices can be partitioned into three sets $\mathcal{V}_{1}$, $\mathcal{V}_{2}$ and $\mathcal{V}_{3}$, then the adjacency matrix $A$ after suitably rearranging its vertices is, 
    \[
    A = \left[ \begin{array}{ccc}
          A_{11} & A_{12} & A_{13} \\
    A_{21} & A_{22} & A_{23} \\
    A_{31} & A_{32} & A_{33}
    \end{array}
    \right]
    \]
    where $A_{ii} \geq 0 \hspace{1.5mm} \forall \hspace{1.5mm} i \in \{1,2,3\} $  and $A_{ij} \leq 0 , i \neq j \hspace{1.5mm} \forall \hspace{1.5mm} i,j \in \{1,2,3\}$. We will not be able to find a diagonal matrix $P$ whose diagonal entries are $+p$ and $-q$ ($p,q>0$), that will satisfy eqn. \eqref{eq:A_condition}. Suppose we make the entries of $A_{12}, A_{21}, A_{23}, A_{32}$ non-negative by similarity transformation using matrix $P$ having $+p$ in the diagonal entries corresponding to the vertices in set $\mathcal{V}_{1}$ and $\mathcal{V}_{3}$ and $-q$ in the diagonal entries corresponding to the vertices in set $\mathcal{V}_{2}$, then the entries corresponding to $A_{13}, A_{31}$ will be non-positive after the similarity transformation.
    Similarly, this proof can be extended to k-partite graphs with $k \geq 3$.
\end{proof}

\begin{theorem}
\label{thm:signed_structurally_unanti-balanced_existence}
Consider a structurally anti-balanced signed graph $\mathcal{G}$, whose corresponding unsigned graph is irreducible and aperiodic. The set $\mathcal{P}$ of matrices, defined in eqn. \eqref{eq:A_condition}, is always an empty set.
\end{theorem}
\begin{proof}
The proof is along the same lines as that of Theorem \ref{thm:signed_structurally_balanced_existence}. We apply the Perron-Frobenius theorem for structurally anti-balanced signed digraphs, which says that the maximum modulus eigenvalue of the adjacency matrix $A$ corresponding to $\mathcal{G}$ is negative, simple, and the largest in magnitude \cite{tian2024spreading}. So, the maximum modulus eigenvalue gives the spectral radius of $A$, but it does not lie in the spectrum of $A$. Further note that for any invertible matrix $S$, the spectrum of $A$ is equal to the spectrum of the matrix $SAS^{-1}$.

Again, we make use of Theorem 3.1 in \cite{valcher2000algebraic} which gives us the necessary conditions for the non-negativity of a matrix. Since the maximum modulus eigenvalue of the matrix $SAS^{-1}$ is not equal to its spectral radius, it cannot make the cone $\mathcal{K}$ invariant. So, $SAS^{-1}$ is not positive for any invertible matrix $S$. Hence, $\mathcal{P}=\{\phi\}$.
\end{proof}

Next, we will design the proposed Laplacian matrix for a special case of a structurally unbalanced graph $\mathcal{G}$, whose subgraph is strongly connected and structurally balanced, and the remaining weakly connected nodes do not pass any information to the strongly connected subgraph. The Adjacency Matrix for this can be written as $A = \begin{bmatrix}
        A_{11} & \mathbb{0} \\
        A_{21} & A_{22}
        \end{bmatrix}$ 
where $A_{11} \in \mathbb{R}^{r\times r}$, $A_{21} \in \mathbb{R}^{r\times n-r}$, $A_{22} \in \mathbb{R}^{n-r\times n-r}$ and $\mathbb{0}$ is a zero matrix of compatible direction. 
Now, the procedure for designing $L_x$ is as given below:
\begin{itemize}
    \item We define a set $\mathcal{P}^{'}$ as the set of all invertible matrices such that,
\begin{equation}
    \label{eq:Set_P_for_struc_unbal}
    \mathcal{P}^{'}:= \{P~|~det(P)\neq 0 \text{ and } P_{1}A_{11}P_{1}^{-1}\geq 0\}
\end{equation}
where $P = \begin{bmatrix}
        P_{1} & \mathbb{0} \\
        \mathbb{0} & P_{2}
        \end{bmatrix}$,
$P_{1} \in \mathbb{R}^{r\times r}$ and $P_{2} \in \mathbb{R}^{n-r\times n-r}$
\item The matrix $P \in \mathcal{P}^{'}$ is then used to find the standard out-degree matrix as, 
\begin{align}
\theta_{z} = diag\left(\mathbb{1}_n^{T}\left(abs\left(PAP^{-1}\right) \right)\right)
\end{align}
where $abs(.)$ gives us the absolute value of the matrix.
\item The modified out-degree matrix $\theta_{x}$ is then given by,
\begin{equation}
    \label{eq:proposed_outdegree_struc_unbalan}
    \theta_{x} =  P^{-1}\theta_{z} P = P^{-1}diag\left(\mathbb{1}_n^{T}\left(abs\left(PAP^{-1}\right) \right)\right) P
\end{equation}
The modified Laplacian matrix can be calculated using eqn. \eqref{eq:linear_system_1} as $L_x=\theta_x-A$.
\end{itemize} 

Now that we have proposed the design of the Laplacian matrix for this special case of structurally unbalanced graphs, a natural question to ask, is the existence of a suitable $P \in \mathcal{P}^{'}$ for this special case of structurally unbalanced graphs. 
\begin{theorem}
    \label{thm:struc_unbaln_existence}
    Consider the special case of structurally unbalanced graph $\mathcal{G}$ whose subgraph is strongly connected and structurally balanced and the remaining weakly connected nodes do not pass any information to the strongly connected subgraph. The set $\mathcal{P}^{'}$ of matrices, defined in eqn. \eqref{eq:Set_P_for_struc_unbal}, is always non-empty.
\end{theorem}
\begin{proof}
    The proof is along the same lines as Theorem \ref{thm:signed_structurally_balanced_existence}
\end{proof}

Now that we have proved the existence of the proposed Laplacian matrix $L_{x}$ for the unsigned graphs, signed structurally balanced graphs, and a special case of structurally unbalanced graphs, in the next section we will show the effect of the Laplacian matrices of these graphs on the evolution of opinion of the agents.

\section{OPINION FORMATION}
\label{sec:Effect_of_proposed_out_degree}
In this section, we will study the evolution of the opinion of the model \eqref{eq:linear_system_1} using the proposed Laplacian matrix $L_{x}$.  
The system defined by eqn. \eqref{eq:linear_system_1} is linear. Hence, its solution is given by,
\begin{equation} 
\label{eq:time_varying_sol}
x(t)=\phi(t,t_0)x_0
\end{equation}
where $\phi(t,t_0) \in\mathbb{R}^{n\times n}$ is the state transition matrix from time $t_0$ to $t\geqslant t_0\geqslant 0$ and $x_0 \in \mathbb{R}^{n}$ is the initial opinions of the agents. Without loss of generality, the initial time $t_0$ is assumed to be $0$ throughout the paper. 

To study the evolution of the opinion states with time, we re-write $-L_{z}$ using its canonical decomposition as $-L_{z}=V_{z}J_{z}W_{z}^{T}$ where $V_{z}$ and $W_{z}$ are the matrices consisting of the right and left eigenvectors of $-L_{z}$, respectively, and $J_{z}$ is the block diagonal Jordan normal form (see section 2.1.2 in \cite{FB-LNS}). 
Since $L_{z}$ is a similarity transformation of the Laplacian matrix $L_{x}$, the spectrum of $-L_{z}$ and $-L_{x}$ is same and is denoted by $\sigma=\{\sigma_1, \sigma_2,\cdots,\sigma_n\}$. 
Then, eqn.  \eqref{eq:time_varying_sol} can be re-written as,
\begin{equation} 
\label{eq:time_var_matrices_sol}
x(t) = P^{-1}V_{z}e^{J_{z}t}W_{z}^{T}Px_0
\end{equation}
The subsequent result aids our understanding of how the proposed Laplacian matrix $L_{x}$ affects opinion formation in unsigned or signed structurally balanced graphs, with a discussion on the stability aspects of the arising opinion evolution.

\begin{theorem}
\label{thm:final_states_general}
For the given proposed Laplacian matrix $L_{x}$ and corresponding invertible matrix $P$ the model \eqref{eq:linear_system_1} admits a stable solution if $PAP^{-1}$ is a non-negative matrix. Let $n_{z}$ be the number of zero eigenvalues of $-L_{x}$, then for this stable case, the steady-state value $\lim_{t\to\infty}x(t)$ can be given as,
\begin{equation} 
\label{eq:fianl_state_general}
x_{f} = P^{-1}\left( \sum_{i=1}^{n_{z}} v_{zi}w_{zi}^{T} \right) Px_{0}
\end{equation}
\end{theorem}
where $v_{zi}$, $w_{zi} \in \mathbb{R}^{n}$ are the right and left eigenvectors corresponding to the $i^{th}$ zero eigenvalue of $-L_{z}$ for $i\in\{1,2,\cdots,n_{z}\}$.

\begin{proof}
Since there exists a proposed Laplacian matrix $L_{x}$, the evolution of eqn. \eqref{eq:linear_system_2} follows the standard Laplacian flow for unsigned graphs and the final opinion states in coordinate system z is stable and is given by, $z_{f} = \left( \sum_{i=1}^{n_{z}} v_{zi}w_{zi}^{T} \right) z_{0}$ and we can use eqn. \eqref{eq:coordinate_change} to arrive at eqn. \eqref{eq:fianl_state_general}. Moreover, the final opinion states in the coordinate system $x$ is also stable since the final opinion states in the coordinate system $z$ is stable.
\end{proof}

\begin{corollary}
For a connected unsigned graph, there is only one zero eigenvalue, and its right and left eigenvectors are $\mathbb{1}_n$ and $\frac{1}{n}\mathbb{1}_n^{T}$ respectively. At
the steady state eqn. \eqref{eq:fianl_state_general} becomes,
\begin{equation} 
\label{eq:fianl_state_unsigned}
x_{f} = \frac{1}{n}P^{-1}\mathbb{1}_{n}\mathbb{1}_n^{T}Px_{0}
\end{equation}
Note that if one of the final states $x_{fi}=0$, then all the other final states are zero for this choice of the diagonal matrix $P$.
Moreover, if we are given the final opinion state, we can compute the required invertible diagonal matrix $P=diag(p_1,p_2,....,p_n) \in \mathbb{R}^{n \times n}$ by,
\begin{equation}
\label{final_states_unsigned_reverse}
k = \frac{1}{n}\sum_{i=1}^{n}p_{i}x_{0i} = p_{1}x_{f1} = p_{2}x_{f2} =...=p_{n}x_{fn}
\end{equation}
where $x_0=(x_{01},x_{02},....,x_{0n}) \in \mathbb{R}^{n}$ is the initial opinion states, $(x_{f1},x_{f2},....,x_{fn}) \in \mathbb{R}^{n}$ is the final opinion states and the initial and final opinion states of the agents follow the constraint $n= \sum_{i=1}^{n}\frac{x_{0i}}{x_{fi}}$

\end{corollary}

The subsequent results aid our understanding of how the proposed Laplacian matrix $L_{x}$ affects the opinion formation in the special case of the structurally unbalanced graph discussed in theorem \ref{thm:struc_unbaln_existence}, with a discussion on the stability aspects of the arising opinion evolution.
\begin{theorem}
    \label{thm:final_states_struc_unbalan}
    For the given proposed Laplacian matrix $L_{x}$ and corresponding invertible matrix $P$ the model \eqref{eq:linear_system_1} admits a stable solution if $P_{1}A_{11}P_{1}^{-1}$ is a non-negative matrix. The Laplacian matrix $-L_{x}$ has one zero eigenvalue and the steady state value $\lim_{t\to\infty}x(t)$ can be given as,
\begin{equation} 
\label{eq:fianl_state_struc_unbalan}
x_{f} =\begin{bmatrix}
        P_{1}^{-1}\mathbb{1}_{r}w_{1}^{T}P_{1}x_{01}  \\
        -P_{2}^{-1}L_{z22}^{-1}L_{z21}\mathbb{1}_{r}w_{1}^{T}P_{1}x_{01}
        \end{bmatrix}
\end{equation}
where $L_{z} = \begin{bmatrix}
        L_{z11} & \mathbb{0} \\
        L_{z21} & L_{z22}
        \end{bmatrix}$ 
, $L_{z11} \in \mathbb{R}^{r\times r}$, $L_{z21} \in \mathbb{R}^{r\times n-r}$, $L_{z22} \in \mathbb{R}^{n-r\times n-r}$ and $v_{1}=\left[\mathbb{1}_{r} \hspace{3mm} -L_{z22}^{-1}L_{z21}\mathbb{1}_{r} \right]$ and $w_{1}=\left[\mathbb{1}_{r} \hspace{3mm} \mathbb{0} \right] \in \mathbb{R}^{n}$ are the right and left eigenvectors corresponding to the zero eigenvalue of $-L_{z}$.
\end{theorem}
\begin{proof}
    The matrix $L_{z22}$ is invertible because its eigenvalues are greater than zero. Since $w_{1}^{T}$ is the left eigen vector of $-L_{z}$, hence  $w_{1}^{T}L_{z11}=0$ and $w_{1}^{T}\mathbb{1}_{r}=1$. The computation of the left and right eigen vectors of the zero eigenvalue for $-L_{z}$ is trivial as $L_{z}$ is a block-triangular matrix. The steady-state value $\lim_{t\to\infty}x(t) = P^{-1}v_{1}w_{1}^{T}Px_{0}$ 
\end{proof}

Using the results discussed in Theorem \ref{thm:final_states_general} and Theorem \ref{thm:final_states_struc_unbalan}, it is possible to obtain the desired opinion clusters for unsigned, signed structurally balanced graphs and the special case of structurally unbalanced graphs which lie in $\mathbb{R}^n$. Clustering of opinions is often a desired outcome as it prevents polarisation. In the next section, we discuss some simulations to illustrate these results.

\section{SIMULATION RESULTS}
\label{sec:Simulation results}
In this section, numerical simulations are presented to validate the theoretical results discussed in the paper. For the subsequent simulations, we consider a signed structurally balanced graph containing three agents with the initial opinion states $x(0)=\left[10, 20, 50\right]^{T}$. 
Furthermore, all the parameters mentioned in this section are in standard units.

\begin{figure}[ht]
\centering
\includegraphics[scale=0.52]{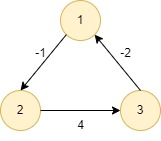}
\caption{Graph topology 1}
\label{Example Graph 1}
\end{figure}

Consider the signed structurally balanced graph shown in Fig. \ref{Example Graph 1}. The adjacency matrix for this graph is  $A_{x}=\left[0, -1, 0; 0, 0, 4; -2, 0, 0\right]$. The spectrum for the adjacency matrix $A$ is $\sigma_{A} = \{-1+1.732i,-1-1.732i,2 \}$. The maximum modulus eigenvalue of this spectrum is $2$ and it is present in the spectrum. Hence, we can find at least one invertible matrix $P$ which satisfies the condition $PAP^{-1}$ is a non-negative matrix.
Now we will see the effect of different invertible matrices $P$ on the evolution of agents whose interactions are given by the graph in Fig. \ref{Example Graph 1}.

\begin{figure}[ht]
\centering \includegraphics[height=0.25\textwidth,width=0.45\textwidth]{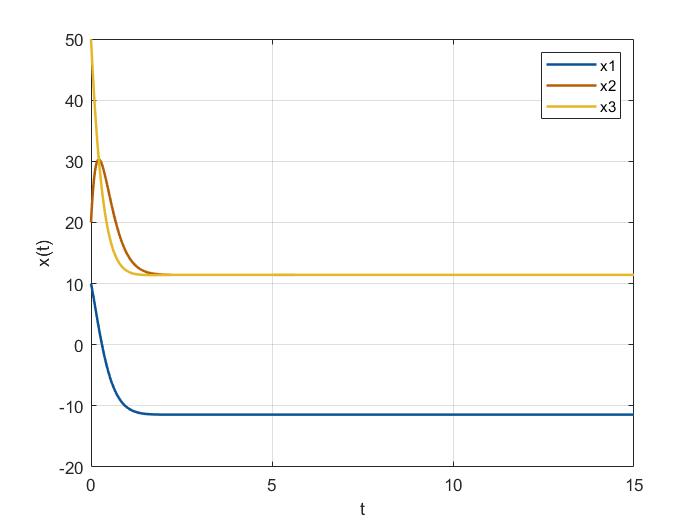}
\caption{Polarisation of opinion states}
\label{polarisation}
\end{figure}

In the first case we will use the invertible matrix $P = \left[2, 0, 0; 0, -2, 0;  0,0, -2\right]$ which satisfies the conditions mentioned in the  Theorem \ref{thm:final_states_general}. The matrix $\theta_{x}$ of the proposed Laplacian matrix for this invertible matrix $P$ is $\theta_{x} = \left[ 1, 0, 0; 0, 4, 0; 0, 0, 2\right]$. The evolution of opinions of the agents that follow this proposed Laplacian matrix is shown in Fig. \ref{polarisation} and the final opinion states of the agents are $x_f = \left[-11.4, 11.4, 11.4 \right]$. These final opinion states are polarized, which is an undesirable outcome. 

\begin{figure}[ht]
\centering
\includegraphics[height=0.25\textwidth,width=0.45\textwidth]{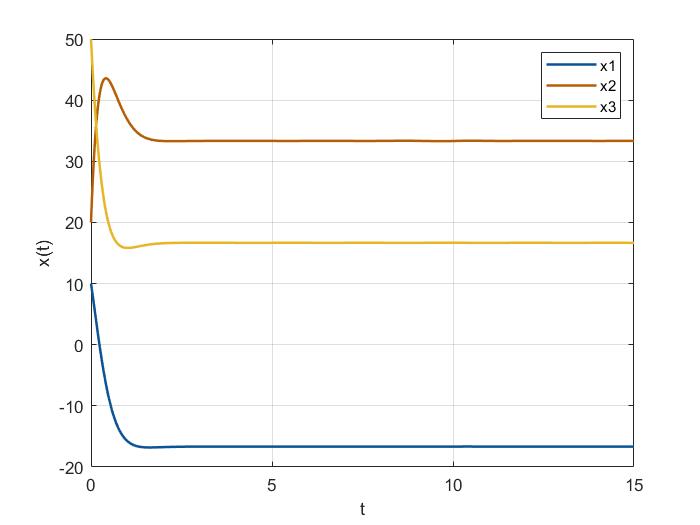}
\caption{Different final opinion states}
\label{clustering}
\end{figure}

In the second case, we will use the invertible matrix $P = \left[-2, -1, 2; -2, 1, -2;  2,1, 2\right]$ which satisfies the conditions mentioned in the  Theorem \ref{thm:final_states_general}. The proposed Laplacian matrix for this invertible matrix $P$ is $\theta_{x} =  \left[ 2, 0, 0; 0, 2, 0; 0, 0, 2\right]$. The evolution of opinions of the agents that follow this proposed Laplacian matrix is shown in Fig. \ref{clustering} and the final opinion states of the agents are $x_f = \left[-16.7, 33.3, 16.7 \right]$.


For the subsequent simulations we consider a special case of the structurally unbalanced graph containing six agents with initial opinion states $x(0) =\left[10, 20, 50, -10, -20, 30\right]^{T}$

\begin{figure}[ht]
\centering
\includegraphics[scale=0.52]{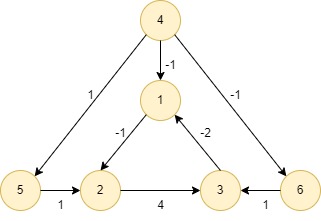}
\caption{Graph topology 2}
\label{Example Graph 2}
\end{figure}

Consider the signed structurally unbalanced graph shown in Fig. \ref{Example Graph 2}. The spectrum for the adjacency matrix $A$ of the graph shown in Fig.\ref{Example Graph 2} is $\sigma_{A} = \{-1+1.732i,-1-1.732i,2,0,0,0 \}$. The maximum modulus eigenvalue of the strongly connected subgraph is $2$ and it is present in the spectrum of the subgraph. Hence, we can find at least one invertible matrix $P$ which satisfies the condition $P_{1}A_{11}P_{1}^{-1}$ is a non-negative matrix.

\begin{figure}[ht]
\centering \includegraphics[height=0.25\textwidth,width=0.45\textwidth]{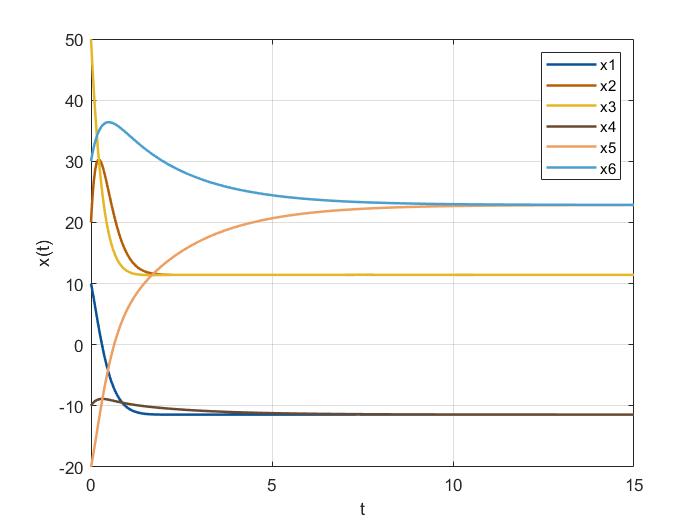}
\caption{Final opinion states for structurally unbalanced graph}
\label{struc_unbalan_final_opinion_states}
\end{figure}

Consider the invertible matrix 
$P = $ [2, 0, 0, 0, 0, 0; 0, -2, 0, 0, 0, 0; 0, 0, -2, 0, 0, 0;  0, 0, 0, 1.2, 0, 0; 0, 0, 0, 0, 1, 0; 0, 0, 0, 0, 0, 1], 
which satisfies the conditions mentioned in the  Theorem \ref{thm:final_states_struc_unbalan}. The matrix $\theta_{x}$ of the proposed Laplacian matrix for this invertible matrix $P$ is 
$\theta_{x}$ = [ 1, 0, 0, 0, 0, 0; 0, 4, 0, 0, 0, 0; 0, 0, 2, 0, 0, 0; 0, 0, 0, 3, 0, 0; 0, 0, 0, 0.5, 0;  0, 0, 0, 0, 0, 0.5]. 
The evolution of opinions of the agents which follow this proposed Laplacian matrix is shown in Fig.\ref{struc_unbalan_final_opinion_states}  and the final opinion states of the agents are $x_f = \left[-11.4, 11.4, 11.4, -11.4, 22.8, 22.8 \right]$. 

\section{DISCUSSION}
\label{sec:Discussion}
In \cite{altafini_polarisation2012}, Altafini considered the strongly connected structurally balanced graphs and proposed a Gauge transform to get polarization of opinions. However, the proposed Laplacian-based model in this paper can be used for structurally balanced graphs whose unsigned counterpart is irreducible and aperiodic, and we were able to show that we can reach the desired final opinion states.

It is shown in \cite{trust_mistrust2016} that the clustering of opinions of the agents occurs using a DeGroot-based model when there is a subnetwork of structurally balanced and globally reachable nodes. However, these clusters are not controlled. So we may not get the desired clusters. However, the proposed laplacian-based model for the special case of structurally unbalanced graphs will give us the desired clusters.

\section{CONCLUSIONS AND FUTURE WORKS}
\label{Sec:conclusion and future works}
%
%
In this paper, we propose the use of a modified Laplacian matrix, which is used to achieve the desired clustering of the final opinion states, which is different from the results obtained for the standard consensus protocol. We stated the conditions for the existence of the proposed Laplacian matrix and showed its existence for unsigned and signed structurally balanced graphs, as well as the special case of structurally unbalanced graphs. We also showed the procedure to design a diagonal invertible matrix, which is used to find the out-of-degree matrix needed for reaching the desired final opinion states. Moreover, we also proved that the proposed Laplacian matrix does not exist for a class of signed anti-balanced graphs and k-partite graphs. We studied the effect of the proposed Laplacian matrix on the opinion evolution of the agents and derived the equations for obtaining the desired final opinion states. Unlike the standard Laplacian-based consensus results, wherein there is no control over the final state vector, in our case, we can achieve any desired opinion clustering. The proposed approach can be used to avoid undesirable outcomes like polarization. We have also presented some numerical simulations to validate the results discussed in the paper.

In the future, we plan to extend the proposed framework to more types of structurally unbalanced graphs. Moreover, we also plan to develop a more general way of finding a non-diagonal invertible matrix that can be used to get the proposed Laplacian matrix.
%
%


The authors gratefully acknowledge the contribution of National Research Organization and reviewers' comments.


References are important to the reader; therefore, each citation must be complete and correct. If at all possible, references should be commonly available publications.


J.G.F. Francis, The QR Transformation I, {\it Comput. J.}, vol. 4, 1961, pp. 265-271.

H. Kwakernaak and R. Sivan, {\it Modern Signals and Systems}, Prentice Hall, Englewood Cliffs, NJ, 1991.

D. Boley and R. Maier, "A Parallel QR Algorithm for the Non-Symmetric eigenvalue Algorithm", {\it in Third SIAM Conference on Applied Linear Algebra}, Madison, WI, 1988, pp. A20.


\bibliographystyle{IEEEtran}
\bibliography{IEEEabrv,main}

\end{document}